\documentclass[aps,superscriptaddress,twocolumn,groupedaddress,pra]{revtex4-1}
\usepackage{graphicx}
\usepackage{xcolor}
\usepackage[colorlinks=true,linkcolor=blue,citecolor=magenta,urlcolor=blue]{hyperref}
\usepackage{physics}
\usepackage{dsfont} 
\usepackage{mathtools}
\usepackage{amssymb}

\begin{document}

\title{Chiral Bell-state transfer via dissipative Liouvillian dynamics}
 \author{Shishir Khandelwal$^{1,2}$, Weijian Chen$^3$, Kater W. Murch$^3$, G\'eraldine Haack$^1$ \vspace{0.2cm}}
\affiliation{$^1$Department of Applied Physics, University of Geneva, 1211 Geneva, Switzerland \\$^2$Physics Department, Lund University, Box 118, 22100 Lund, Sweden\\$^3$Department of Physics, Washington University, St. Louis, Missouri 63130}


\date{\today}

\begin{abstract}
Chiral state transfer along closed loops in the vicinity of an exceptional point is one of the many counter-intuitive observations in non-Hermitian physics. The application of this property beyond proof-of-principle in quantum physics, is an open question. In this work, we demonstrate chiral state conversion between singlet and triplet Bell states through fully-quantum Liouvillian dynamics. Crucially, we demonstrate that this property can be used for the chiral production of Bell states from separable states with a high fidelity and for a large range of parameters. 
Additionally, we show that the removal of quantum jumps from the dynamics through postselection can result in near-perfect Bell states from initially separable states.  Our work presents the first application of chiral state transfer in quantum information processing and demonstrates a novel way to control entangled states by means of dissipation engineering.
 
\end{abstract}

\maketitle

\textit{Introduction.---} In recent years, Exceptional Points (EPs) in non-Hermitian systems have seen a surging interest, for example, in sensing \cite{Chen2017,Zhang2019, Yu2020, Wiersig2020, DeCarlo2022, Wong2023,Larson2023}, topological properties \cite{ Mailybaev2005,Bergholtz2021, Ding2022, Liu2021, Abbasi2022, Okuma2023,Arkhipov2023} and recently in the quantum control of dynamics \cite{Khandelwal2021,Kumar2022a,zhang2022,Zhou2023}. Open quantum systems are a natural platform to explore EPs and associated effects as their evolution is characteristically non-Hermitian \cite{Minganti2019}. Chiral state transfer (CST) along closed trajectories in the vicinity of an EP is a well-known property of non-Hermitian physics, where eigenstates can be adiabatically switched and the final state is solely determined by the orientation of the trajectories \cite{Berry2011}. While this effect was first discussed for classical and semiclassical systems \cite{Dembowski2003,Uzdin2011,Doppler2016,xu2016,Hassan2017,Nasari2022}, its applications to quantum settings are only now being discovered. Theoretical works \cite{Kumar2021,Sun2023} have been accompanied by successful experiments with superconducting circuits \cite{Chen2021,Chen2022} and single ions \cite{Bu2023}. These results offer a pathway for quantum state control by utilising dissipation as a resource. However, presently, they do not necessarily involve genuinely quantum phenomena, such as the creation of quantum correlations like entanglement, or offer an advantage in quantum settings. \par
In this work, we show for the first time, that it is possible to create highly entangled states by means of CST between two Bell states. 
We consider a minimal model of two interacting qubits, put in an out-of-equilibrium situation by coupling to thermal environments. This model has been used to demonstrate the presence of entanglement in the steady-state regime under autonomous dissipative dynamics only \cite{BohrBrask2015,Tacchino2018, Khandelwal2020, Prech2023,Khandelwal2024} and has recently been investigated in the context of EPs \cite{Khandelwal2021}. We demonstrate that slow trajectories in the parameter space of this system can result in CST between two Bell states. Importantly, this property can be utilized to create highly entangled states from arbitrary initial states. We further demonstrate that the transfer fidelity and entanglement can be increased by means of postselection, at the cost of reduced success rate \cite{Minganti2020,Kumar2021}. Finally, we demonstrate that our results hold for a wide range of parameters, including trajectories not necessarily encircling EPs \cite{Nasari2022, Abbasi2022}.  \par
\textit{Model and trajectory in parameter space--}
\begin{figure}
    \centering
    \includegraphics[width=1\columnwidth]{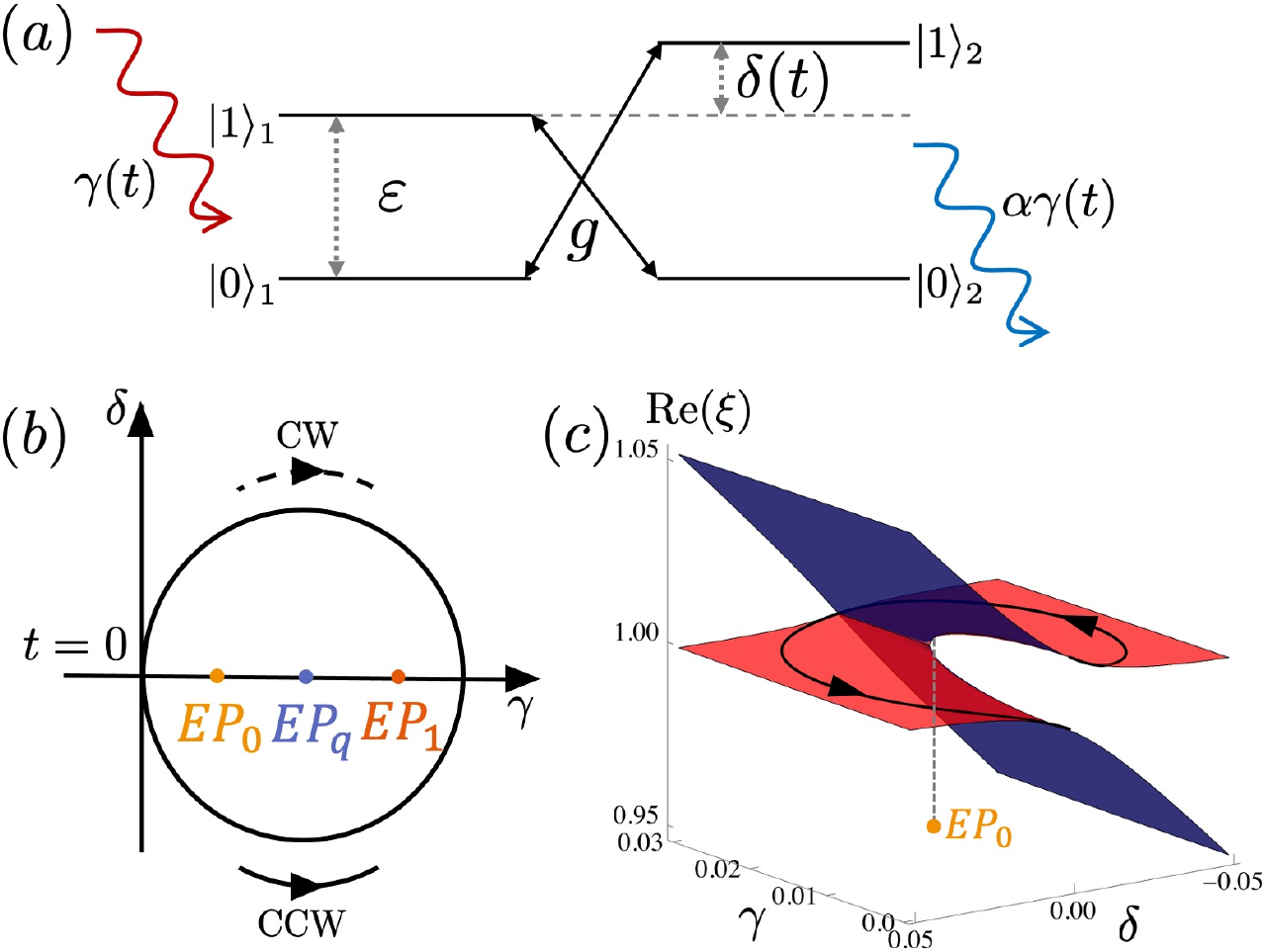}
    \caption{(a) 
    Two interacting qubits subject to gain and loss. (b) Clockwise (CW) and Counterclockwise (CCW) trajectories in the space of $\delta$ and $\gamma$, and depictions of EPs for different levels of postselection of quantum jumps, $q=0$ (complete postselection), $q=1$ (no postselection) and $0<q<1$ (partial postselection). (c) Riemann sheets corresponding to the NHH eigenvalues involved in the EP and a CCW trajectory showing state transfer. The more decaying branch of the eigenvalues (i.e., with more negative imaginary part) is depicted in red.}
    \label{fig:setup}
\end{figure}
We consider a system of interacting qubits, depicted in Fig. \ref{fig:setup} (a), with transition energies $\varepsilon_1=\varepsilon$, $\varepsilon_2=\varepsilon+\delta$. The full Hamiltonian takes the form $H_0=\sum_{j=1,2}\varepsilon_j\sigma_+^{(j)}\sigma_-^{(j)}+g (\sigma_+^{(1)}\sigma_-^{(2)}+\sigma_-^{(1)}\sigma_+^{(2)} )$ ($j=1,2$), where $\sigma^{(j)}_\pm$ are the raising and lowering operators of qubit $j$ and $g$ is the interaction strength. Each qubit couples to its own fermionic environment 
with coupling strength $\gamma_j$. Under the assumption $g,\gamma_j, \delta \ll\varepsilon$, the Markovian evolution of the two qubits can be expressed in the following Lindblad form (with $\hbar = k_B =1$) \cite{Breuer2007,Hofer2017,Potts2021}, 
\begin{align}
\label{eq:Lind}
    \dot\rho = \mathcal{L} \rho = -i\left[ H_\text{eff},\rho\right]_\dagger+ \sum_{j=1,2} \gamma_j^+ \mathcal{J}_+^{(j)} \rho   + \gamma_j^- \mathcal{J}_-^{(j)} \rho,
\end{align}
where $[a,b]_\dagger\coloneqq ab -b^\dagger a^\dagger$ and the effective non-Hermitian Hamiltonian (NHH) $H_\text{eff} \coloneqq H_0 - (i/2)\sum_j \gamma_j^-\sigma_+^{(j)} \sigma_-^{(j)} +\gamma_j^+ \sigma_-^{(j)} \sigma_+^{(j)}$. The superoperators $\mathcal J_\pm^{(j)} \rho \coloneqq \sigma_\pm^{(j)} \rho \sigma_\mp^{(j)}$ represent quantum jump terms.
The corresponding excitation and de-excitation rates $\gamma_j^+=\gamma_j n_j$ and $\gamma_j^-=\gamma_j (1-n_j)$ are determined by the Fermi-Dirac distribution, $n_j = (e^{\beta_j(\varepsilon_1+\varepsilon_2)/2}+1)^{-1}$ with inverse temperature $\beta_j$  of reservoir $j$. We refer to these rates as ``gain" and ``loss", respectively. The NHH induces coherent non-unitary loss, while quantum jumps represent continuous monitoring by the environment \cite{Wiseman}. It is possible to interpolate between purely NHH and fully quantum dynamics by using a hybrid-Liouvillian framework \cite{Minganti2020}, with a quantum jump parameter, $q\in [0,1]$,

\begin{align}
\label{eq:hybrid_Lind}
\mathcal{L}_{[q]} \rho =  -i\left[ H_\text{eff},\rho \right]_\dagger 
+ q \sum_{j=1,2} \gamma_{j}^+ \mathcal{J}_+^{(j)} \rho +  \gamma_j^- \mathcal{J}_-^{(j)} \rho.
\end{align}The case $q=1$ corresponds to full Lindblad dynamics, while $q=0$ and $0<q<1$ to complete and partial postselection of quantum jumps, respectively. The spectra of $H_{\text{eff}}$ and $\mathcal L_{[q]}$ and corresponding EPs are discussed in the Supplementary Material (SM) Secs. I and II. Importantly, the NHH has a second-order EP involving eigenvectors which become Bell states $\ket{\Psi^\pm}=\left( \ket{10}\pm\ket{01}\right)/\sqrt{2}$ when the system is decoupled from the reservoirs (conversely, the eigenvectors $\ket{\Psi^\pm}$ of $H_0$ are involved in an EP in the presence of dissipation). By judicious choice of a parameter trajectory, it is expected that these Bell states play a central role in CST. This property is demonstrated on the Riemann sheets corresponding to the eigenvalues of $H_{\text{eff}}$ depicted in Fig. \ref{fig:setup} (c), and forms the basis of the upcoming analysis. \par

\begin{figure}
    \centering
    \includegraphics[width=0.85\columnwidth]{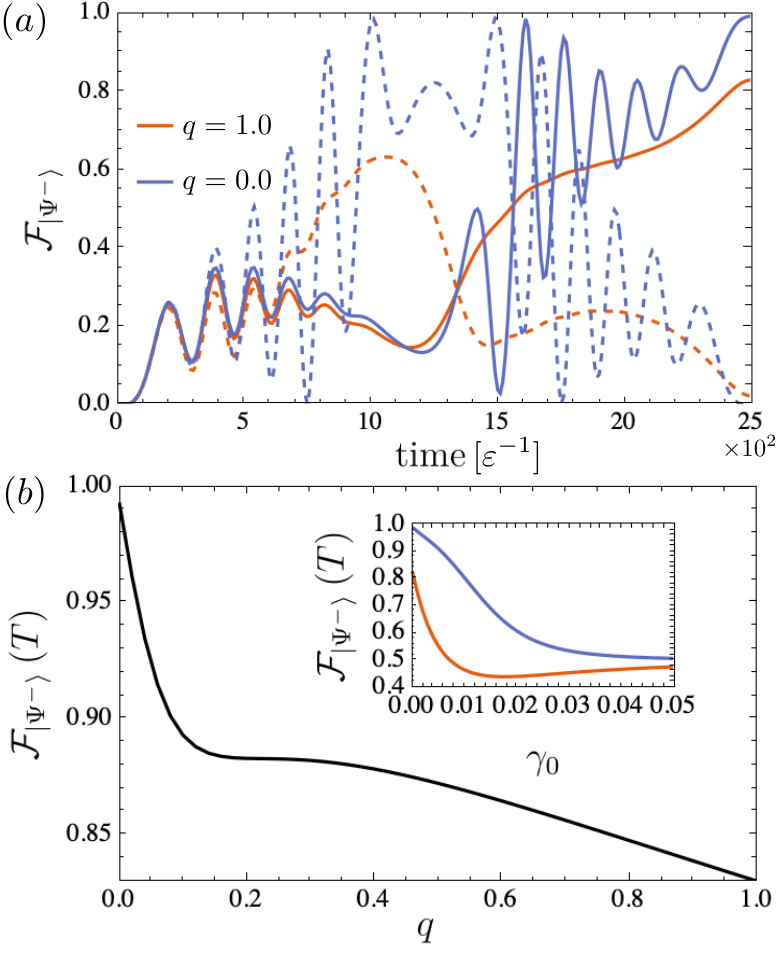}
    \caption{
    	 (a) Fidelity of $\rho(t)$ with the 
    Bell state $\ket{\Psi^-}$ as a function of time, for CW (dashed) and CCW (solid) trajectories. (b) The final fidelity (at time $T$) as a function of the quantum jump parameter $q$, shown only for the CCW trajectory. The inset shows the fidelity as a function of $\gamma_0$, which sets the origin of the trajectory, for $q=0$ (blue) and $q=1$ (red). The parameters are $\varepsilon=1$, $\Delta\delta/\varepsilon=0.04$, $\gamma_0=0$, $\Delta\gamma/\varepsilon=0.008$, $g/\varepsilon=0.01$, $\alpha=1.2$, $\beta_1\to-\infty$, $\beta_2\to\infty$ and $T\varepsilon=2500$. }
    \label{fig:CSC1}
\end{figure}
We set $\gamma_1 \coloneqq \gamma$ and $\gamma_2 \coloneqq\alpha \gamma_1$ with $\alpha=\gamma_2/\gamma_1$ and choose a closed trajectory in the space of $\gamma$ and $\delta$ (Fig.~\ref{fig:setup} (b)). We pick the following form of periodic driving of the parameters, 
\begin{align}
\label{eq:traj}\hspace*{-0.2cm}
\delta(t) =\pm \Delta\delta \sin\left(\frac{2\pi t}{T}\right),\,\,
\gamma(t) = \gamma_0 + \Delta\gamma \sin^2\left(\frac{\pi t}{T} \right),
\end{align}where $\gamma_0$ sets the origin, $\Delta \delta$ and $\Delta \gamma$ are the amplitudes for $\delta(t), \gamma(t)$ respectively and $T$ is the driving period. The ``$+$" and ``$-$" signs correspond to Clockwise (CW) and Counterclockwise (CCW) trajectories, respectively. By taking an appropriately large $\Delta\gamma$, the trajectory can be made to encircle the EPs. For any degree of postselection, $0\leq q\leq1$, the unnormalized state evolves according to, $\rho(t) = \mathcal T \text{exp}\left[ \int_0^t \mathcal{L}_{[q]}(t') dt' \right] \rho(0)$, where $\mathcal T$ denotes time ordering. However, for $0\leq q<1$, the dynamics are not trace-preserving. The following numerical results for $0\leq q<1$ have been calculated by discretizing the above time-ordered exponential and numerically renormalizing the state at every time-step.

\textit{Chiral Bell-state transfer.--} We now turn to the core of this Letter: CST in the vicinity of EPs. We first investigate the case $\gamma_0=0$, in which the initial and final points (i.e., $t=0$ and $t=T$) of the trajectory are $\gamma=\delta=0$. In this case, at the initial and final points of the trajectory, $H_0$ and $H_\text{eff}$ have the same spectrum and eigenvectors, $\{ \ket{11}, \ket{\Psi^+}, \ket{\Psi^-}, \ket{00}\}$. Moreover, at $t=0$, the eigenmatrices of $\mathcal L$ can be constructed from the eigenvectors of $H_0$ and $H_{\text{eff}}$ \cite{Minganti2019}. This equivalence between Hamiltonian, NHH, and Liouvillian dynamics at the origin of the trajectory is essential for chiral Bell-state transfer.  For a setup with fermionic reservoirs, we choose the inverse temperatures $\beta_1 \to -\infty$ and $\beta_2\to\infty$. This corresponds to perfect population inversion in reservoir 1 ($n_1=1$) and to initialization in the lowest energy level ($n_2=0$) for reservoir 2, leading to the gain and loss situation shown in Fig. \ref{fig:setup} (a). As we will see later, this corresponds to the optimal setup for chiral Bell-state transfer. It also has a connection to PT symmetry in the setup (see SM Sec. IV and Ref. \cite{Huber2020}). We characterize the state along the trajectory with its fidelity with respect to one of the Bell states $\mathcal{F}_{\ket{\Psi^\pm}}(t) \coloneqq \Tr \{\ketbra{\Psi^\pm}{\Psi^\pm} \rho(t)\}$.
\par
When $q=0$,  $H_{\text{eff}}$ dictates the dynamics and CST is expected with near-perfect fidelity due to the conservation of purity (see SM Sec. III). At the origin of the trajectory, the two states involved in the EP are $\ket{\Psi^\pm}$. 
Starting in $\ket{\Psi^+}$, the system either switches to $\ket{\Psi^-}$ or ends up again in $\ket{\Psi^+}$, depending on the orientation of the trajectory. In Fig.~\ref{fig:CSC1} (a) (blue curves), we show this effect starting with $\ket{\Psi^+}$, which ends up at time $t=T$ nearly perfectly in the $\ket{\Psi^-}$ state for a CCW trajectory (solid) or comes back to $\ket{\Psi^+}$ for a CW trajectory (dashed). Therefore, switching between the states is chiral in nature. As shown in the SM Sec. V, for $q=0$, we find that maximal transfer fidelity can be achieved simply by taking sufficiently slow trajectories, i.e., CST is an adiabatic property.  \par
Without complete postselection ($q \neq 0$), maximal transfer fidelity cannot be reached as Liouvillian dynamics do not preserve purity \cite{Sergi2013}. 
 The case $q=1$, which corresponds to full Liouvillian dynamics, is shown in red in Fig.~\ref{fig:CSC1} (a) reaching a final fidelity $\mathcal{F}_{\ket{\Psi^-}}= 0.83$. We note that this is not an upper bound, and a higher fidelity can be achieved by suitably altering the parameters. The chirality of the state $\ket{\Psi^-}$ can similarly be verified; switching to $\ket{\Psi^+}$ is observed for a CW trajectory, while the state returns to $\ket{\Psi^-}$ for a CCW one, with the transfer fidelity remaining the same.   
In Fig.~\ref{fig:CSC1} (b), we show that there is a monotonic decrease in transfer fidelity with increasing $q$ (decreasing postselection). This corresponds to recent observations \cite{Kumar2021,Chen2021} and can be traced to the loss of purity with quantum jumps. Crucially, the inset in Fig. \ref{fig:CSC1} (b) shows  a drastic fidelity loss for trajectories with origins far from $\gamma_0=0$. When the trajectory origin is chosen away from $\gamma_0=0$, Bell states are not the eigenstates of the NHH at the start and end of the trajectory. This means that state transfer occurs between some other states, which may be far from Bell states. This highlights the importance of the spectra of $H_0$, $H_{\text{eff}}$ and $\mathcal L$ being equivalent at the origin. Our analysis shows that CST is a property of the eigenstates of $H_{\text{eff}}$. While $\mathcal L_{[q]}$ (for $0<q\leq1$) shows it to a large extent, the fidelity is lowered due to loss of purity induced by quantum jumps.

\begin{figure}
    \centering
    \includegraphics[width=0.85\columnwidth]{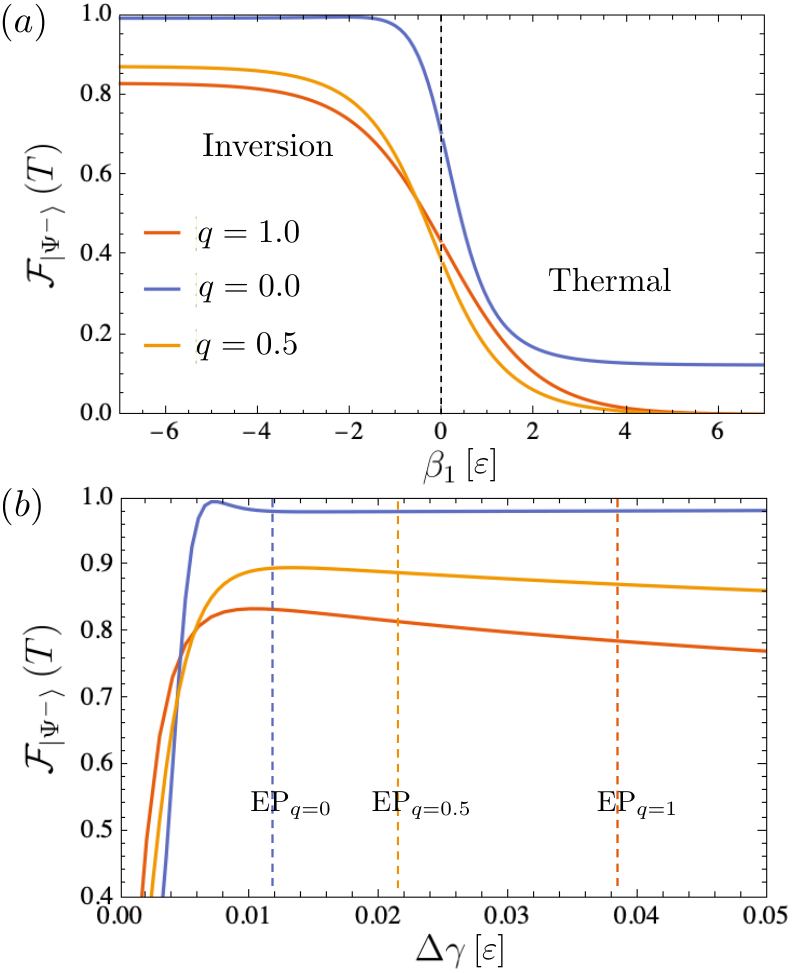}
    \caption{(a) 
    	The final fidelity $\mathcal{F}_{\ket{\Psi^-}}(T)$ as a function of $\beta_1$, with $\beta_2\to\infty$. (b) The final fidelity as a function of the amplitude $\Delta\gamma$. The trajectories are EP-encircling on the right of each corresponding dashed line. The other parameters are taken from Fig. \ref{fig:CSC1}.
    }\label{fig:temp}
\end{figure}For $q=1$, the behaviour can be analytically understood through the spectrum of the one-cycle evolution superoperator $\mathcal{P}(T) = \mathcal{T} \exp \left[ \int_0^T \mathcal{L}(t') dt' \right]$ \cite{Hartmann2017}. It has eigenvalues which are either 1 or lie within the unit circle. For long driving time periods $T$, the fixed point of the superoperator is reached, which corresponds to the unique eigenmatrix with eigenvalue 1. In the time-independent case, this eigenmatrix with eigenvalue 1 is equivalent to the eigenmatrix with eigenvalue 0 of the time-independent Liouvillian $\mathcal L$, corresponding to the usual steady state. This fixed point is independent of the initial state and only depends on the 
system and driving parameters. This can be understood within the framework of Floquet theory \cite{Hartmann2017,Schnell2020}, which can also be useful to look at the general case, $q\in[0,1]$, with a corresponding $\mathcal P_{[q]}(T)$. While slow trajectories (i.e., with large $T$) are required for higher fidelities (see SM Sec. V), 
adiabatic trajectories will drive the system to its instantaneous steady state at every point in the
trajectory. Driving too slow may also lead to a loss of chirality \cite{Sun2023}. We expect further insights may come from slow-driving perturbation theory, by calculating corrections to adiabatic evolution \cite{Cavina2017}.

 We now extend our predictions to the case where transport is not unidirectional, i.e., without perfect control of dissipation. We let $\beta_2\to\infty$, implying perfect loss at qubit 2, and calculate the final fidelity as a function of $\beta_1$ ($\beta_1<0$ implying population inversion). The absence of complete population inversion induces a decrease in the fidelity as shown in Fig.~\ref{fig:temp} (a). Optimal fidelity is achieved for perfect population inversion, $\beta_1 \to -\infty$, independently of $q$. Moreover, for thermal environments ($\beta_1>0$),
  there is a drastic reduction in fidelity, the maximum is achieved for low $\beta_1$ ($\beta_1 \to 0$ or $n_1 \to 1/2$). Our analysis demonstrates that high transfer fidelity requires $n_1>1/2$, which corresponds to population inversion. It can further be verified that simultaneously, a low temperature for reservoir 2 is required for a high fidelity.\par
 Finally, let us comment on the role of EPs in our predictions. It has recently been observed in some semiclassical settings that encircling EPs is not necessary for CST \cite{Hassan2017,Nasari2022}. We find that encircling EPs in our setup is not necessary to achieve chiral Bell-state transfer. We illustrate this in  Fig.~\ref{fig:temp} (b) for different values of $q$, by varying the radius $\Delta \gamma$ of the trajectory along $\gamma$ (see Eq. \eqref{eq:traj}). The left side of each dashed line corresponds to trajectories not encircling the corresponding EP, while the right side corresponds to encircling trajectories. The plot showcases the robustness of CST to changes in $\Delta\gamma$. Interestingly, we find that the maximum fidelity is obtained for trajectories not encircling the EPs. Whether there is a fundamental principle underlying this observation is an interesting question beyond the scope of this work.
\begin{figure}
    \centering
    \includegraphics[width=0.85\columnwidth]{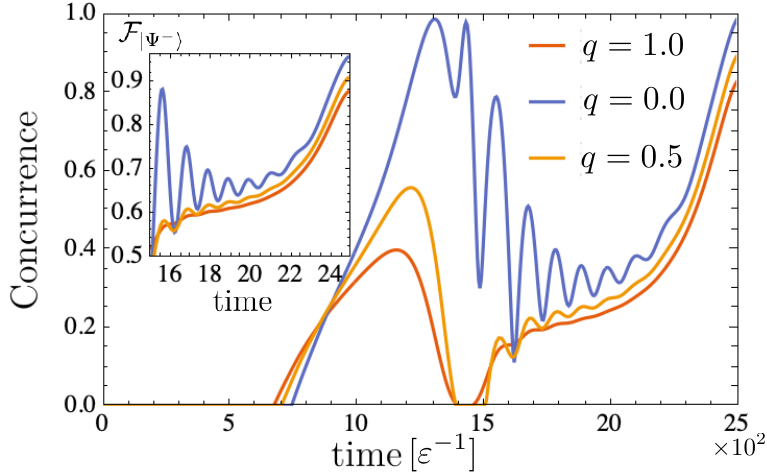}
    \caption{Concurrence as a function of time with $\rho(0)=\mathds 1/4$, for various $q$. The inset shows the corresponding fidelity with $\ket{\Psi^-}$.  The parameters are $\varepsilon=1$, $\Delta\delta/\varepsilon=0.06$, $\gamma_0=0$, $\Delta\gamma/\varepsilon=0.01$, $g/\varepsilon=0.01$, $\alpha=1.2$, $\beta_1\to-\infty$, $\beta_2\to\infty$ and $T\varepsilon=2500$. The trajectory is CCW.}
    \label{fig:CST_ent}
\end{figure}

\textit{Chiral production of Bell states.--} We now exploit chiral Bell-state transfer to demonstrate the generation of maximal two-qubit entanglement starting from any separable initial state. We consider the two qubits to be initially in a maximally mixed state, $\rho(0) = \mathds 1/4$ and take the same parameter driving as discussed above. Apart from the fidelity,  we characterize the amount of entanglement through the concurrence $\mathcal C$ \cite{Wootters1998}. For a density matrices $\rho$ involved in our scheme, its expression simplifies to $\mathcal C(\rho)\coloneqq2\,\text{max}\{0,\lvert c\rvert-\sqrt{p_{11}p_{00}}\}$, with the populations $p_{11} = \bra{11}\rho\ket{11}$ and $p_{00}=\bra{00} \rho \ket{00}$, and coherence $c =\bra{01}\rho \ket{10}$; $\mathcal{C}=0$ for a separable state, and $\mathcal{C}=1$ for a maximally entangled state. In Fig. \ref{fig:CST_ent}, the concurrence increases from 0 at $t=0$, to its maximum at time $t=T$, where the system is driven close to the $\ket{\Psi^-}$ state. Importantly, for any $q$, a high concurrence can be obtained; specifically for $q=1$, $\mathcal C>0.83$. Although this is not an upper bound, it far exceeds the highest possible concurrence, $\mathcal C_{\text{max}}^{(\text{aut})}\approx0.31$ \cite{BohrBrask2022,Prech2023}, possible with the two-qubit system being operated autonomously (i.e., in the absence of driving or external control). Crucially, the production of Bell states has an associated chirality; for a CCW state, the system is driven to the $\ket{\Psi^-}$ state, while for a CW trajectory, to the $\ket{\Psi^+}$ state. Although the final state is independent of the initial state (for sufficiently large $T$), it is dependent on the parameters of system. As the inset shows, a higher fidelity $\mathcal F\sim 0.9$ is obtained for the $q=1$ case than in Fig. \ref{fig:CSC1}.

\textit{Experimental scope.--} We anticipate that an experimental implementation is readily achievable in the circuit QED platform \cite{blais21}, by utilizing a superconducting device consisting of two transmon circuits \cite{koch07} that interact via a resonator-mediated coupling. The two transmons have individual readout resonators \cite{gaikwad_2024} that allow us to introduce the respective thermal baths. Positive and negative temperature fermionic baths can be replaced with engineered bosonic baths harnessing cavity assisted bath engineering \cite{Murch2012}, where population inversion can be achieved through off-resonant driving of the qubit and associated cavity. Here, the qubit  is driven with a detuning $\Delta$ and resonant Rabi frequency $\Omega_0$. The cavity is driven with a detuning $\Delta' = \pm \sqrt{\Delta + \Omega_0}$. For positive (blue detuned) cavity drive, the qubit is pumped to the excited state, the purity of the inversion is limited by the ratio of the cavity assisted bath engineering rate (ultimately limited by the cavity decay rate $\kappa =1.3\  \mathrm{rad.}/\mu\mathrm{s}$), to the coherence times of the qubit. In this setup, coherence times are approximately $T_1, T_2* \simeq 30 \ \mu$s. Hence we expect $n =0.98$.   The limit $q=1$ (Lindblad dynamics)  can be attained by harnessing the lowest two energy levels of the transmon \cite{Chen2021,Chen2022}, while $q=0$ can be accessed by utilizing its higher manifold of states coupled with postselection \cite{Naghiloo2019}. Postselection, however, comes with the cost of reduced success probability for the protocol. The fidelity of the $q=1$ limit can be deterministically improved by preventing transitions to the $\ket{11}$ state (see SM Sec. VI). This can be achieved in superconducting platform through the Zeno effect to decouple this state from the dynamics \cite{Harrington2022, Blumenthal2022}.

\textit{Discussion.--} We have theoretically demonstrated Liouvillian CST involving Bell states in a system of two dissipative qubits. This property allows for the chiral production of near-Bell states starting with any separable initial state, breaking the bounds for autonomous dissipative entanglement production. The results hold for a large range of parameters, operating in the vicinity of an EP, without the necessity of encircling it. \par
Our results have implications beyond simple two-qubit models, and present a recipe for quantum state control through controlled dissipation and clever eigenstate engineering. For example, by judicious choice of many-qubit Hamiltonians and dissipation, our results suggest that CST, and by extension, entanglement production, can be seen for genuinely multipartite entangled states, like the W or GHZ  state \cite{Dur2000}.
\par

\textit{Acknowledgements.--} We thank Soumya S. Kumar for contributions to an early stage of this project and Parveen Kumar for useful correspondence. SK and GH acknowledge support from the SNSF through the starting grant PRIMA PR00P2\textunderscore179748, as well as the NCCR SwissMAP for supporting short scientific visits.  WC and KM acknowledge support from NSF Grant No. PHY-1752844 (CAREER), and the Air Force Office of Scientific Research (AFOSR) Multidisciplinary University Research Initiative (MURI) Award on Programmable systems with non-Hermitian quantum dynamics (Grant No. FA9550-21-1- 0202).

\bibliography{References.bib}

\clearpage
\newpage

\onecolumngrid

\section*{Supplementary Material}

\section{Full spectrum}
\label{app:spectrum}

\begin{figure*}[h]
	\centering    \includegraphics[width=1\textwidth]{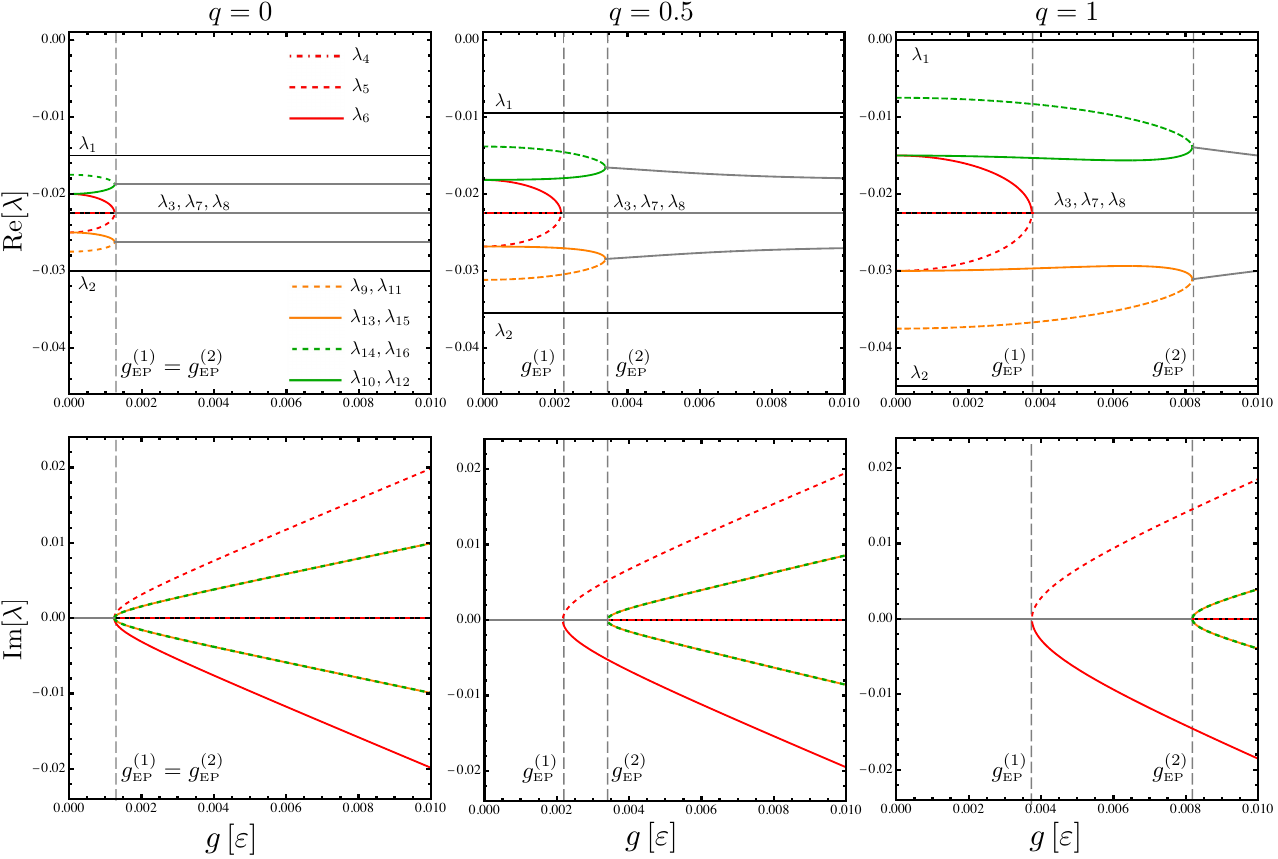}
	\caption{Real and imaginary parts of the eigenvalues of $\mathcal L_q$ as functions of the inter-qubit coupling $g$. The eigenvalues that merge at EPs are depicted in the same color. The parameters are $\gamma_1^-/\varepsilon = 0.02$, $\gamma_1^+/\varepsilon=0.01$, $\gamma_2^-/\varepsilon = 0.01$ and $\gamma_2^+/\varepsilon = 0.005$. The energy scale, $\varepsilon=1$. The eigenvalues $\lambda_{1-6}$ correspond to the reduced Liouvillian discussed in Supp. Mat. Sec. \ref{app:EPs} C.}
	\label{fig:spectrum}
\end{figure*}

\noindent In this section, we discuss the full spectrum of the time-independent hybrid Liouvillian, $\mathcal L_{[q]}$. Figure \ref{fig:spectrum} shows all 16 eigenvalues, real and imaginary parts in the upper and lower rows, respectively, for 3 different values of $q$. For $q=1$, as expected, there is a unique steady state, which corresponds to the eigenmatrix with eigenvalue $0$. Moreover, in all three cases, we find that real part of all non-zero eigenvalues is negative. This means that the system is pushed into its least decaying eigenmatrix (corresponding to eigenvalue $\lambda_1$) at long times.

There are several Exceptional Points (EPs) in the Liouvillian (of second, third and fourth order). However, only one third-order EP plays a significant role in the dynamics of the two-qubit system towards its steady state, involving eigenvalues $\lambda_4,\,\lambda_5$ 
and $\lambda_6$ in red. As discussed in detail in Ref.~\cite{Khandelwal2021}, in the $q=1$ case, this EP marks critical dynamics towards the steady state. From Fig.~\ref{fig:spectrum}, we see that plots from eigenvalues are parabolic at $q=0$ and merge at several sets of EPs which correspond to the same value $g=g_{\text{\tiny{EP}}}^{(1)}=g_{\text{\tiny{EP}}}^{(2)}$. When quantum jumps are present ($0<q\leq1$), however, these plots become deformed and two sets of EPs emerge, which correspond to different values $g_{\text{\tiny{EP}}}^{(1)}\neq g_{\text{\tiny{EP}}}^{(2)}$.\par
We focus on a reduced subspace and only the eigenvalues $\lambda_{1-6}$, as discussed in detail in the next section.

\section{Eigenvalues, eigenvectors and presence of EPs}
\label{app:EPs}

\subsection{Hamiltonian}

\noindent The eigenvalues of the two-qubit Hamiltonian $H_0$ take the following form,
\begin{equation}
	\theta_1 = 0, \,\, \theta_{2,3} =\varepsilon + \delta/2 \pm \sqrt{ g^2 + (\delta/2)^2}\,\, \text{and}\,\, \theta_4 = 2\varepsilon +\delta,
\end{equation}with the eigenvectors,
\begin{equation}
	\begin{aligned}
		&\textbf{v}_1 = (1,0,0,0 )^T = \ket{11}\,\\
		&\textbf{v}_{2,3} = \left(0, \delta/(2g) \pm \sqrt{1+ (\delta/(2g))^2}, 1, 0 \right)^T \\
		&\textbf{v}_4 = (0,0,0, 1 )^T = \ket{00}.
	\end{aligned}
\end{equation}For $\delta=0$, the eigenvectors $\textbf v_{2,3}$ reduce to the Bell states $\ket{\Psi^\pm}=\left(\ket{10}\pm\ket{01}\right)\sqrt{2}$.

\subsection{Effective Hamiltonian}

\noindent The eigenvalues of the effective Hamiltonian are given below,
\begin{equation}
	\begin{aligned}
		\xi_1 = -\frac{i}{2} \Gamma^+, \,\,\xi_{2,3}=\frac{1}{4}\left( -i\Gamma \pm \eta_0 +4\varepsilon+2\delta\right),\,\,\text{and}\,\,\,\xi_4 = -\frac{i}{2} \Gamma^-+2\varepsilon + \delta.
	\end{aligned}
\end{equation}We define $\Gamma_i\coloneqq\gamma_i^+ + \gamma_i^-$, $\Gamma^\pm\coloneqq \gamma_1^\pm+\gamma_2^\pm$, $\Gamma\coloneqq \Gamma^+ + \Gamma^-$, $\tilde\Gamma_i\coloneqq\gamma_i^--\gamma_i^+ $ and $\eta_0\coloneqq i\sqrt{\left( \tilde \Gamma_1 -\tilde \Gamma_2 -2i\delta\right)^2-16g^2}$. The corresponding eigenvectors are
\begin{equation}
	\begin{aligned}\label{eq:eigsH}
		&\boldsymbol x_1= \ket{00}, \\
		&\boldsymbol x_{2,3} = \left(0,\frac{-2\delta-i\left( \tilde \Gamma_1 - \tilde\Gamma_2\right) \pm\eta_0}{4g},1,0 \right)^T, \\
		&\boldsymbol x_4 = \ket{11}.
	\end{aligned}
\end{equation}$H_{\text{eff}}$ has an EP at $\delta=0$ and $\eta_0 = 0$, where eigenvalues and eigenvectors 2 and 3 merge. At the origin of the trajectory, $\delta = \gamma =0$, the eigenvectors $\boldsymbol x_{2,3}$ correspond to the $\ket{\Psi^\pm}$ eigenvectors of the Hamiltonian $H$. Moreover, at the origin, the 16 eigenmatrices of $\mathcal L$ are $\boldsymbol x_i\boldsymbol x_j^T$ ($i,j=1,...,4$). This is why the two Bell states play a special role when considering the parameter trajectories depicted in Fig. 1 (b) in the main text, and chiral state transfer is seen between them. This is further depicted in Fig. 1 (c) in the main text, where we show the Riemann sheets corresponding to eigenvalues $\xi_{2,3}$, along with evolution along a parameter trajectory.

\subsection{Hybrid Liouvillian}

To investigate Liouvillian dynamics, we restrict to the steady-state subspace of the two-qubits, which is most relevant \cite{Khandelwal2021}. The hybrid Liouvillian in matrix form, $L_{[q]}$ (in the basis $\{\ketbra{11}{11},\ketbra{10}{10},\ketbra{10}{01},\ketbra{01}{10},\ketbra{01}{01},\ketbra{00}{00}\}$), takes the following form,

\begin{align}
	L_{[q]} = \left(
	\begin{array}{cccccc}
		-\gamma_1^--\gamma_2^- & \gamma_2^+ q & 0 & 0 & \gamma_1^+ q & 0 \\
		\gamma_2^- q & -\gamma_1^--\gamma_2^+ & i g & -i g & 0 & \gamma_1^+ q \\
		0 & i g & \frac{1}{2} (2 i \delta -\Gamma) & 0 & -i g & 0 \\
		0 & -i g & 0 & \frac{1}{2}  (-2i\delta-\Gamma) & i g & 0 \\
		\gamma_1^- q & 0 & -i g & i g & -\gamma_1^+-\gamma_2^- & \gamma_2^+ q \\
		0 & \gamma_1^- q & 0 & 0 & \gamma_2^- q & -\gamma_1^+-\gamma_2^+ \\
	\end{array}
	\right).
\end{align}The eigenvalues and eigenvectors of the Liouvillian are in general complicated. We write the eigenvalues here for the case with $\delta=0$, which is sufficient to see several different EPs. The eigenvalues take the following form,
\begin{equation}
	\begin{aligned}
		\lambda^q_{1,2} = \frac{1}{2}\left( -\Gamma \pm \eta_q^{(1)}\right), \quad\lambda^q_{3,4} = -\frac{\Gamma}{2}\quad \text{and}\quad  \lambda^q_{5,6} = \frac{1}{2}\left( -\Gamma \pm \eta_q^{(2)}\right),
	\end{aligned}
\end{equation}where,
\begin{equation}
	\begin{aligned}
		&\eta_q^{(1,2)} \coloneqq \sqrt{-8g^2+\sum_{j=1,2}\left( \left(\gamma_j^- - \gamma_j^+\right)^2+4q^2\gamma_j^-\gamma_j^+\right) \pm 2\beta_q} \\
		&\beta_q = \sqrt{16 g^4 + 8 g^2 [\gamma_1^- \gamma_2^- + \gamma_1^+ \gamma_2^+ - (\gamma_1^+ \gamma_2^- + \gamma_1^- \gamma_2^+) (1-q^2)] + \big(\Gamma_1^2 - 4(1-q^2) \gamma_1^- \gamma_1^+\big)\big(\Gamma_2^2 - 4(1-q^2) \gamma_2^- \gamma_2^+\big)}.
	\end{aligned}
\end{equation}In the case $q=1$, the above simplify to $\eta_1^{(1)}=\Gamma$ and $\eta_1^{(2)}=\sqrt{\left(\Gamma_1 - \Gamma_2 \right)^2-16g^2}$. This gives us $\lambda_1^{1}=0$ (which corresponds to the steady state). In the case $q=0$, we have $\eta_0^{(1)}=\sqrt{\left(\tilde\Gamma_1-\tilde\Gamma_2 \right)^2-16g^2}$ and $\eta_0^{(2)} = \left(\tilde\Gamma_1+\tilde\Gamma_2 \right)$. \par
To summarize the EPs discussed in the main text:
\begin{enumerate}
	\item{For general $0<q<1$, there is an EP at $\eta_q^{(2)}=0$.}
	\item{For $q=1$, the EP lies at $\eta_1^{(2)}=0$ \cite{Khandelwal2021}.}
	\item{For $q=0$, the EP lies at $\eta_0= i\eta_0^{(1)} =0 $, where $\eta_0$ was defined above. This is identical to the EP of the NHH.}
\end{enumerate}

\section{Purity}
\label{app:purity}
\noindent We have noted in the main text that the hybrid-Lindblad equation,
\begin{align}
	\dot\rho(t) =\mathcal L_{[q]}\rho(t) = -i\left[H_{\text{eff}},\rho(t) \right]_\dagger + q\sum_j L_j\rho(t)L_j^\dagger   ,
\end{align}is not trace preserving for $0\leq q<1$. The probability loss can be quantified by,
\begin{align}
	\frac{d}{dt}\text{Tr}\left\{\rho(t)\right\} = -\left(1-q \right)\sum_{j}\text{Tr}\left(L_j^\dagger L_j\rho\left(t\right) \right),
\end{align}which makes it clear that the equation is trace preserving for $q=1$. Moreover, the rate is proportional to $(1-q)$, which means smaller the quantum-jump parameter and lesser the effect of quantum jumps, larger the loss in probability for any given time. The hybrid-Lindblad equation can be renormalized by introducing the following nonlinear term, $ (1-q)\sum_j\text{Tr}\left\{ L_j^\dagger L_j\rho(t) \right\}\rho(t)$. The addition of this term makes the overall (normalized) evolution equation,
\begin{equation}\label{eq:renorm}
	\begin{aligned}
		\dot{\rho}(t) = 
		\mathcal L_{[q]}\rho(t) + (1-q)\sum_{j}\text{Tr}\left\{ L_j^\dagger L_j\rho(t) \right\}\rho(t).
	\end{aligned}
\end{equation}It can be verified that the solution to Eq. \eqref{eq:renorm} is given by the normalized state, $\rho(t) = e^{\mathcal L_{[q]} t}\rho\left(0\right)/\text{Tr}\left\{e^{\mathcal L_{[q]} t}\rho\left(0\right) \right\}$. Therefore, the above equation could have equivalently been used for numerical calculations shown in the main text and would require no renormalization by hand.\par
With the renormalized equation \eqref{eq:renorm}, we can now discuss the evolution of the purity.  In general, Lindblad evolution has a ``mixing" effect on the state of the system, meaning that a pure state does not evolve into another pure state. This can be seen clearly by considering the rate of change of the purity of the state, $d\text{Tr}\left\{\rho^2 \right\}/dt \equiv \dot{ P}\left(t\right) = 2\text{Tr}\left\{\rho\left( t\right)\dot\rho\left( t\right) \right\}$. Using this with Equation \eqref{eq:renorm}, we obtain,
\begin{equation}\label{eq:purity}
	\begin{aligned}
		\frac{1}{2}\dot{ P}\left(t \right) = \sum_j  q \left[\text{Tr}\left\{L_j^\dagger\rho L_j\rho\right\} - \text{Tr}\left\{L_j^\dagger L_j\rho\right\} P\left(t\right) \right] 
		-\text{Tr}\left\{L_j^\dagger L_j\rho^2\right\} + \text{Tr}\left\{L_j^\dagger L_j\rho\right\}P\left(t\right).
	\end{aligned}
\end{equation}The last two terms in the RHS of the above equation sum to zero, when $P(t)=1$ (or equivalently, $\rho=\rho^2$), i.e., when the state is pure. This means that if $q=0$ and $ P(t)=1$, then $\dot{ P}(t)=0$ and the purity of the state of the system remains conserved. In other words, the NHH evolves pure states into pure states. We note that while NHH evolution preserves the purity of pure states, it does not preserve the purity of mixed states; this distinguishes NHH evolution from usual unitary evolution in quantum mechanics \cite{Sergi2013}.  It is therefore, clear that coherent loss with the NHH alone, does not mix an initially pure state. Instead, the mixing effect comes from the quantum jump terms. This is the reason why the Lindblad equation cannot possibly lead to perfect chiral Bell-state conversion and necessarily leads to some purity loss, which is what we observe in the main text. Eq. \eqref{eq:purity} allows for the quantification of the loss of purity for (hybrid) Lindblad evolution.

\section{Parity-Time symmetry}
\label{app:PT}
To discuss PT-symmetry in the two-qubit system, we use the definition from Ref. \cite{Huber2020}. PT symmetry may be consistently defined  for a dissipative bipartite quantum system by defining the Parity operator to exchange the subsystems and the time operator as a Hermitian conjugation of Lindblad operators (to effectively interchange gain and loss), and acting the PT operator separately on the Hamiltonian and Lindblad operators. Mathematically, PT symmetry then corresponds to
\begin{equation}\label{eq:PT}
	\mathcal L[\mathbb{PT}(H),\mathbb{PT}(L^{(1)}_j),\mathbb{PT}(L^{(2)}_j)] = \mathcal L[H,L_j^{(1)},L_j^{(2)}],
\end{equation}
where $L_j^{(1,2)}$ are Lindblad jump operators acting on system 1 or 2 as defined. For instance, for a single qubit, they would correspond to $\sigma_\pm$. For the dissipative two-qubit system, we may define the parity operator to exchange the qubits,
\begin{align}
	\mathbb{P} : 1 \leftrightarrow 2,
\end{align}
and the time operator to exchange the Lindblad operators, effectively inverting the role of gain  and loss,
\begin{align}
	\mathbb{T}: \sigma_{\pm}^{{(j)}} \leftrightarrow \sigma_{\mp}^{(j)}.
\end{align}
We obtain the following conditions for the system to be PT-symmetric,
\begin{eqnarray}
	\label{eq:condPT}
	&& \delta = -2\varepsilon \\
	&& \gamma_\pm^{(1)} = \gamma_\mp^{(2)}\,,
\end{eqnarray}
where $\gamma_\pm^{(j)}$ are the effective rates for the jumps $\sigma_\pm^{(j)}$. The first condition corresponds to having one qubit transition of energy $\varepsilon$ and the other of $-\varepsilon$, or to have both transitions of energy 0 (the latter implicitly corresponds to the model considered in \cite{Huber2020}). The other conditions amount to imposing that the loss from one qubit is equal to the gain into the other, and vice-versa. For fermionic reservoirs, as considered in this work, this can be expressed as,
\begin{eqnarray}
	&& \gamma_{1} n_{1} = \gamma_{2} (1- n_{2})  \\
	&&\gamma_{1} (1-n_{1}) = \gamma_{2} n_{2}.
\end{eqnarray} 
The above two conditions impose $n_1 + n_2 = 1$, a condition that is physically realized, for instance, with population inversion on one side and zero temperature on the other. This is exactly the case considered in the main text to achieve chiral state transfer with the highest fidelity.

\section{Role of trajectory origin and driving period $T$}
\begin{figure}
	\centering
	\includegraphics[width=\textwidth]{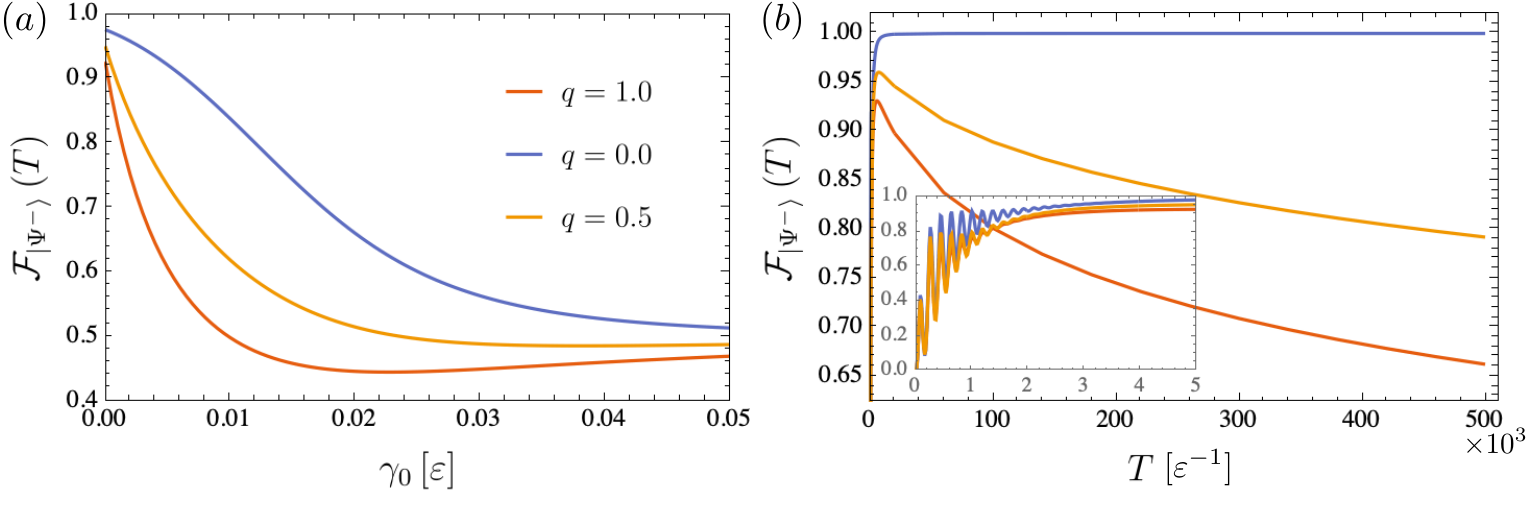}
	\caption{(a) Final state fidelity $\mathcal F_{\ket{\Psi^-}}(T)$ as a function of the trajectory origin $\gamma_0$ with $T\varepsilon=4\times10^3$. (b) Final state fidelity $\mathcal F_{\ket{\Psi^-}}(T)$ with the driving time $T$ with $\gamma_0=0$. The inset zooms in on a smaller time scale. The other parameters are $\varepsilon=1$, $\gamma/\varepsilon=0.01$, $\alpha=1.2$, $g/\varepsilon=0.01$, $n_1=1$ and $n_2=0$.}
	\label{fig:fig2}
\end{figure}

As we have emphasized throughout the main text, chiral state transfer is a property of the non-Hermitian Hamiltonian, in which eigenstates involved in an EP (given in Eq. \eqref{eq:eigsH}) can be switched by taking slow trajectories in parameter space. So if we start in the eigenstate $\boldsymbol{x}_2(0)$, it can become $\boldsymbol x_3(T)=\boldsymbol x_3(0)$, if an appropriate closed trajectory is taken. Therefore, if at the start of the trajectory, we have $H_{\text{eff}}=H$ (which happens when the system is decoupled from the reservoirs, $\gamma=0$), we should see state transfer between eigenstates of the Hamiltonian, i.e., the Bell states. This is why the condition $\gamma_0 =\gamma(0) =0$ is necessary to see chiral state transfer between Bell states. Moving away from $\gamma_0=0$, the Bell states are no longer eigenstates of the $H_{\text{eff}}$ at the start and end of the trajectory. This is expected to lead to a loss of fidelity or entanglement. This is demonstrated in Fig. \ref{fig:fig2} (a), where a clear drop in the transfer fidelity away from $\gamma_0=0$ is observed for all three values of $q$ considered.

In Fig. \ref{fig:fig2} (b), we further investigate the role of the driving time $T$. Under NHH evolution, perfect transfer fidelities (or equivalently, maximally entangled states) can be obtained simply by taking a slow enough trajectory. This is depicted in Fig. \ref{fig:fig2} (b) in the $q=0$ curve, which shows $\mathcal F_{\ket{\Psi^-}}\to1$ for long $T$. However, this is not the case with Liouvillian and hybrid-Liouvillian evolution. Here, we find that taking a trajectory that is too slow leads to a loss of fidelity and eventually, loss of chirality. This can be seen in the $q=1$ and $q=0.5$ curves in Fig. \ref{fig:fig2} (b). Here, the fidelity reaches a maximum and then drops monotonically. We note that this observation, specifically with $q=1$, has been previously made in \cite{Sun2023} for a different setup.

\section{Enhancing transfer fidelity by suppressing double occupation}
\noindent As explained in the main text, the concurrence for the states considered here takes the form $\mathcal C=2\,\text{max}\{0,\lvert c\rvert - \sqrt{p_{11}p_{00}}\}$, where $c$ is the coherence corresponding to the element $\ketbra{10}{01}$ and $p_{11}$ ($p_{00}$) is the probability corresponding to the state $\ket{11}$ ($\ket{00}$). Evidently, by suppressing or damping the probability of double occupation ($p_{11}$), the concurrence can be increased and a state closer to the Bell state $\ket{\Psi^-}$ (or $\ket{\Psi^+}$) can be obtained \cite{BohrBrask2022,Prech2023,Khandelwal2024}. For fermions, such a situation can occur through natural Coulomb repulsion, which can completely suppress the probability of double occupation when the fermions are charged (for example, in electronic quantum dots \cite{ReimannRMP2002}). For engineered bosonic reservoirs, such a situation can be realized by decoupling the $\ket{11}$ state from the dynamics through the quantum Zeno effect (for example, in superconducting qubits \cite{Harrington2022, Blumenthal2022}).
The overall consequence is that the transition $\ket{01}\to\ket{11}$ is suppressed. We refer to the corresponding transition rate as $\tilde \gamma$. In the case of no suppression (i.e., the usual evolution described in the main text), this rate is simply the usual rate $\gamma_1^+=\gamma_1(1-n_1)=\gamma_1$. On the other hand, in the case of complete suppression, this rate is $\tilde\gamma=0$.

\begin{figure}
	\centering
	\includegraphics[width=0.5\textwidth]{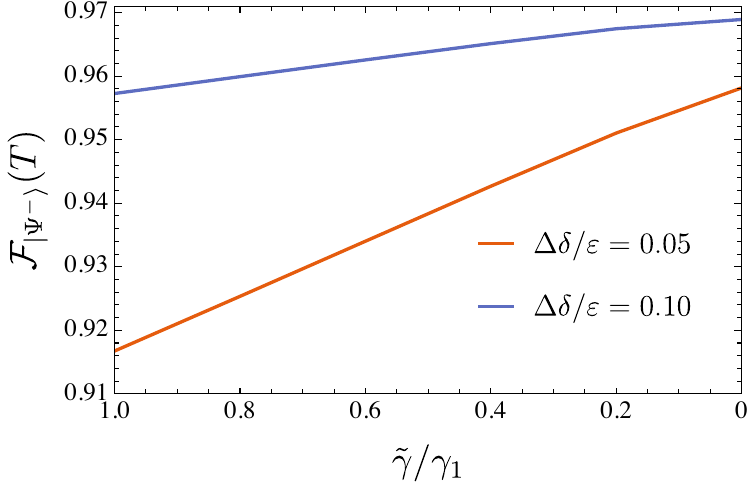}
	\caption{Final state fidelity as a function of $\tilde\gamma/\gamma_1$, with the initial state $\ket{\Psi^+}$ and a CCW trajectory. $\tilde\gamma/\gamma_1 = 1$ corresponds to no suppression of the $\ket{11}$ state (i.e., the usual evolution described in the main text) and $\tilde\gamma/\gamma_1=0$ corresponds to the complete suppression of $\ket{11}$ from the dynamics. The parameters are $\varepsilon=1$, $T\varepsilon=10\times10^3$, $\gamma_0=0$, $\gamma/\varepsilon=0.4\times10^{-2}$, $\alpha=0.8$, $g/\varepsilon=0.8\times10^{-2}$, $n_1=1$, $n_2=0$ and $q=1$.  }
	\label{fig:withU}
\end{figure}

We demonstrate this idea in Fig. \ref{fig:withU}, where we plot the fidelity as a function of the ratio $\tilde\gamma/\gamma_1$. We find that when this ratio decreases (i.e. when double occupation is suppressed), the transfer fidelity is enhanced. The plots are obtained without any postselection - for the considered parameters, we are able to achieve almost 97\% fidelity. These figures are representative and not optimal; obtaining a bound on fidelity will require optimization over all parameters of the system and driving.

\end{document}